# Impact of charge carrier injection on single-chain photophysics of conjugated polymers

Felix J. Hofmann, Jan Vogelsang*, John M. Lupton

Institut für Experimentelle und Angewandte Physik, Universität Regensburg,

Universitätsstrasse 31, 93053 Regensburg, Germany

*jan.vogelsang@physik.uni-regensburg.de

**Abstract**

Charges in conjugated polymer materials have a strong impact on the photophysics and their interaction with the primary excited state species has to be taken into account in understanding device properties. Here, we employ single-molecule spectroscopy to unravel the influence of charges on several photoluminescence (PL) observables. The charges are injected either stochastically by a photochemical process, or deterministically in a hole-injection sandwich device configuration. We find that upon charge injection, besides a blue-shift of the PL emission and a shortening of the PL lifetime due to quenching and blocking of the lowest-energy chromophores, the non-classical photon arrival time distribution of the multichromophoric chain is modified towards a more classical distribution. Surprisingly, the fidelity of photon antibunching deteriorates upon charging, whereas one would actually expect the number of chromophores to be reduced. A qualitative model is presented to explain the observed PL changes. The results are of interest to developing a microscopic understanding of the intrinsic charge-exciton quenching interaction in devices.

The possibility to unravel static and dynamic heterogeneity from the spectroscopy of single isolated molecules is a unique strength of single-molecule spectroscopy (SMS),[1-3] but also implies inherent disadvantages. One of these limitations is the fact that the molecule must be fluorescent. For this reason, usually only non-charged molecules are studied, because charges and subsequent irreversible chemical reactions lead to a strong quenching of the PL and bleaching of the chromophore.[4-6] As a consequence, it is not straightforward to transfer the knowledge obtained from SMS on conjugated polymers as used, e.g., for organic light-emitting diodes to a final device. Charges and their electric fields have a strong impact on the final device characteristics, e.g. exciton diffusion,[7] PL quantum yield[8-10] and PL emission wavelength.[4,11]

The random formation of charges on a single conjugated polymer chain is responsible for the highly dynamic changes in PL intensity, i.e. fluorescence blinking, and PL lifetime.[1,12] The PL exhibits single-step blinking events that have been attributed to efficient intramolecular excitation energy transfer towards a positive charge residing within a trap state.[1,13,14] Subsequent chemical reactions, which can lead to irreversible scission of the polymer chain, also lead to a diminished PL.[15] In this case, the reduction in PL intensity has a different reason, e.g. a lower absorption cross-section or a decreased transition oscillator strength. These different PL quenching mechanisms can only be distinguished with regards to reversibility: photochemical damaging of states is irreversible whereas photobleaching by photoinduced charging is reversible.[16] In combination with the random nature of charge generation, i.e. radical formation, it can be very difficult to investigate the PL characteristics of a single charged conjugated polymer chain. Additionally, the low PL intensity during such quenching events poses a further hurdle to extracting enough information from the remaining photon stream.

The remaining photon stream in the fluorescence of the single chain contains valuable information. For example, the photons during a deterministic (electrically triggered photoinduced)[17] or stochastic (merely photoinduced)[4] fluorescence quenching event were used to localize the origin of the remaining PL with nanometer precision by employing a super-resolution microscopy approach.[18,19] It could be shown that the charge is always situated on the same part of the conjugated polymer chain upon subsequent repeated charge injection events.[18] In another investigation, it was demonstrated recently that a comparison of the photon statistics during a random PL quenching event with unquenched PL from a single conjugated polymer chain can be used to extract information on the nature of the charge trap states, e.g. whether a single charge trap exists or whether multiple quenching sites arise in the multichromophoric system.[20] Generally, the photon statistics of conjugated polymers can be used to estimate the number of independently emitting chromophores under the assumption that the chromophores exhibit the same brightness.[21,22] Intuitively, one would expect that the number of independently emitting chromophores simply decreases during a PL quenching event and that the photon statistics therefore change towards a more non-classical distribution (i.e. improved photon antibunching).[23] Together with the possibility to inject charges at will in a device configuration, the fascinating possibility arises to switch the photon statistics between classical and non-classical behavior for the PL of a single conjugated polymer chain. However, we will demonstrate and explain in the following that the photon statistics respond in the opposite way to this intuitive expectation.

A controlled ansatz to inject charges deterministically into single chains holds promise to measure and evaluate the remaining PL in more detail. Based on a charge injection device, developed by Bolinger *et al.*,[24] we performed SMS measurements on poly[2-methoxy-5-(2-ethylhexyl-oxy)-1,4-phenylene-vinylene] (MEH-PPV) and extracted several observables simultaneously during repeated quenching of the PL: (i) PL intensity, (ii) PL lifetime, (iii) PL emission wavelength and (iv) PL photon statistics. We find that upon charge injection the PL intensity drops within several seconds towards a plateau. The reduced PL intensity is accompanied by a shortening of the PL lifetime, $\tau_{PL}$, by ~ 10 % and a blue-shift of the peak emission wavelength, $\lambda_{Peak}$, by ~ 3nm. Most interestingly, the photon statistics change from clearly non-classical characteristics (photon antibunching) of the non-quenched PL to more classical behavior during PL quenching. We compared the PL characteristics (i-iv) during such a controlled charge injection with random photoinduced blinking events of single MEH-PPV chains and observed the same trends, concluding that deterministic (triggered) and stochastic (random) charge injection are subject to the same physics. The change in photon statistics cannot be described in the simple picture that a certain set of chromophores contributing equally to the overall PL is simply switched off by injecting a charge.

Controlled charge injection was achieved in a device structure sketched in Fig. 1(a). First, the insulating SiO$_2$ layer was deposited onto a patterned indium tin oxide (ITO) coated coverslip by plasma-enhanced chemical vapor deposition (PECVD). Second, a solution of single-molecule concentration MEH-PPV and the inert transparent matrix poly(methyl-methacrylate) (PMMA) in toluene was spin cast for 2 minutes at 2000rpm, resulting in a 25nm layer of PMMA hosting isolated sample molecules. Due to the low thickness of the PMMA film, most molecules are at the surface and can therefore be contacted electrically. This contacting was realized by three additional layers, which were subsequently evaporated onto the PMMA layer. A 200nm thick layer of the organic semiconductor 4,4'-Bis(N-carbazolyl)-1,1'-biphenyl (CBP) was used for positive charge injection into MEH-PPV due to its low-lying HOMO level. Next, 25nm of N,N'-Bis(3-methylphenyl)-N,N'-diphenylbenzidine (TPD) act as an intermediate layer, followed by 100nm of gold as the top electrode. All evaporation steps were conducted at deposition rates below 1Å/s in a high vacuum evaporation chamber and all samples were encapsulated before removal from the glove box to prevent photo-oxidation.

Transients of the PL intensity were recorded using a confocal microscopy setup described in detail in Ref. [25]. For excitation of MEH-PPV, a pulsed laser beam with a wavelength of 485nm and 80MHz repetition rate was used at 200W/cm². The PL emission was recorded by two avalanche photodiodes (APDs) configured in a Hanbury-Brown Twiss geometry, i.e. the PL emission was split by a 50/50 beam splitter and photon coincidence on the two detectors was measured. Fig. 1(b) shows an example of a PL transient of a single MEH-PPV chain within the device structure recorded over 20 seconds. By applying a positive bias of 1V to the gold electrode a time 5s into the measurement, positive charges (holes) are repeatedly injected into and extracted out of single chains. The PL decreases slowly over 2s towards a plateau, suggesting that an equilibrium is reached between charge injection and extraction.[24] Upon switching the bias to -3V after 10s, the holes are extracted and the PL signal recovers instantaneously. This switching process is reproducible over many cycles facilitating the spectroscopic investigation of the quenched state in conjugated polymers by controlled injection of charges.

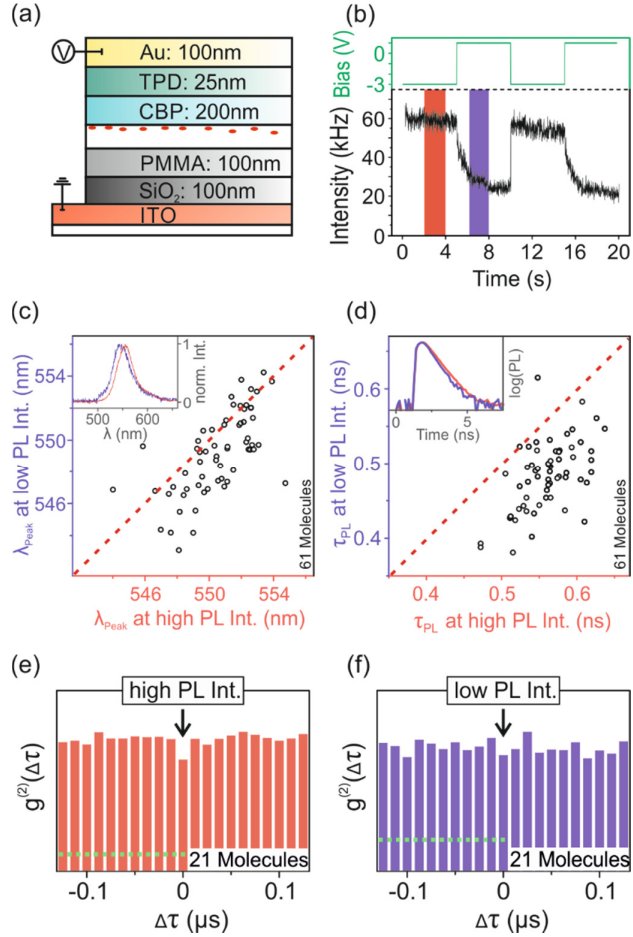

FIG. 1. Charge injection device for electromodulated single-molecule spectroscopy of MEH-PPV. (a) Structure of the charge injection device. The red dots indicate single MEH-PPV chains embedded in a thin PMMA layer. (b) Measurement of PL intensity under voltage modulation for a single MEH-PPV chain, in which the PL transient is shown in black and the applied bias is shown above, in green. The regions of the measurement from which the peak emission wavelength, $\lambda_{Peak}$, and the PL lifetime, $\tau_{PL}$, were extracted are marked in red (high PL intensity) and blue (low PL intensity), respectively. (c, d) Scatterplot of $\lambda_{Peak}$ and $\tau_{PL}$ for the high PL intensity (red) and low PL intensity (blue). The insets show two examples of spectra and PL decay curves (red (blue) for the high (low) PL intensity) from which $\lambda_{peak}$ and $\tau_{PL}$ were extracted. (e, f) Accumulated second-order cross-correlation photon coincidence histograms for 21 molecules in the high (e) and low intensity (f) state. The green dotted line in (e) and (f)

corresponds to the expected g$^{(2)}$(0) value for a perfect single-photon emitter based on the measured signal-to-background ratio.

We recorded the PL spectrum for high and low intensity intervals for 61 single MEH-PPV chains. The inset of Fig. 1(c) shows the spectra of the same MEH-PPV chain under -3V bias (red spectrum) and +1V bias (blue spectrum), which were recorded during the corresponding time intervals marked in the transient in Fig. 1(b). The peak emission wavelength, $\lambda_{Peak}$, was extracted by fitting the 0-0 transition with a Gaussian function. For the example shown in the inset of Fig. 1(c) a blue-shift of $\lambda_{Peak}$ by ~8nm can be observed upon hole injection. The scatter plot in Fig. 1(c) displays the extracted $\lambda_{Peak}$ values for the unquenched (x-axis) and quenched (y-axis) state of each chain, respectively. The red dashed line corresponds to a line through the origin with slope one, which demonstrates that most (70 %) of the single chains exhibit a blue-shift in PL emission upon hole injection with an average blue-shift of ~2.5nm. However, a small fraction (14 %) of the chains exhibit a slight red-shift upon charging with an average of ~1.6nm. The rest of the molecules reveal no spectral shift or the shift is not reproducible over at least two charge injection cycles.

Second, the PL lifetime, $\tau_{PL}$, was extracted by time-correlated single photon counting (TCSPC) of the same molecules and simultaneously to the measurement of $\lambda_{Peak}$. The inset of Fig. 1(d) shows representative PL lifetime decays of one chain under -3V (red) and +1V bias (blue). A mono-exponential fit was applied to extract $\tau_{PL}$, which is plotted in Fig. 1(d) as a scatter plot for the neutral (x-axis) and charged (y-axis) state, respectively. Over 95% of the single chains exhibit a decrease in $\tau_{PL}$ upon hole injection. While the mean of $\tau_{PL}$ is ~570ps in the neutral state, it decreases to ~480ps in the charged state.

Using a Hanbury-Brown Twiss setup allows the evaluation of the fluorescence photon statistics with a standard start-stop photon counting experiment. The histograms in Fig. 1(e) and (f) display the number of photon coincidences at both APDs with a time delay of up to ten excitation pulses between the APDs. A coincidence at time delay *nT*, where *T* is the period of the pulsed laser, is composed of a stop-photon detected at $APD_2$ *nT* excitation pulses after a start-photon was detected at $APD_1$, or *vice versa*. The accumulated correlation histogram in Fig. 1(e) of 21 molecules in the neutral state (-3V bias) reveals a dip of ~15% at zero time delay (marked by a black arrow) between both APDs. An accumulated correlation histogram of the same 21 molecules in the charged state (+1V bias) is presented in Fig. 1(f) with no discernible dip at zero time delay. Following Ref. 22 we expect that the uncorrelated background signal gives rise to an increase in the value of $g^{(2)}(0)$. For this reason, we calculated the expected value of $g^{(2)}(0)$ for the case of a perfect single-photon emitter and the signal to background ratio measured here (see green dotted line in Fig. 1(e), (f)). It is important to note that a lower or higher bias, respectively, has no significant impact on observables reported here. However, the close proximity of MEH-PPV to the hole injection layer (CBP) can impact $\lambda_{Peak}$, $\tau_{PL}$ and the photon statistics. We therefore compared these observables with the case of random charging by photoinduced radical formation and fabricated standard samples used for single-molecule spectroscopy.[1,26]

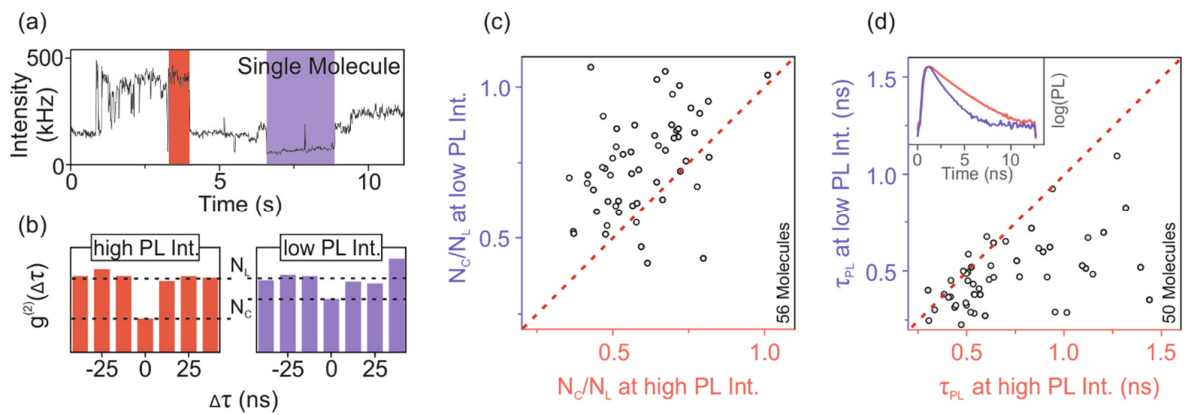

FIG. 2. Stochastic photoinduced charging events in single-molecule spectroscopy of MEH-PPV. (a) Typical PL intensity transient with the areas marked in red and blue from which the photon statistics shown in (b) were derived. In contrast to Fig. 1, sufficient statistics can already be extracted from a single chain. (c) The ratio between the (central) values at zero time delay between both detectors, $N_C$, and non-zero (lateral) time delays, $N_L$, were extracted for each single chain at high PL intensity (red) and low PL intensity (blue) and summarized in a scatter plot. (d) The PL lifetime, $\tau_{PL}$, was extracted for the same time regions and summarized in a scatter plot. The inset shows typical PL decays for high (red curve) and low (blue curve) PL intensity.

Single chains of MEH-PPV were embedded in a host matrix, PMMA, by dynamically spin-coating a toluene/PMMA/MEH-PPV solution onto a cleaned cover glass. To facilitate photoinduced charging and jumps in fluorescence intensity, the samples were not sealed in the nitrogen glove box but instead held in a chamber constantly purged with nitrogen, giving rise to a nitrogen atmosphere slightly contaminated with ambient oxygen. Fig. 2(a) shows a PL transient of a single MEH-PPV chain excited at 485nm with 200W/cm² excitation intensity. The typical reversible blinking behavior on the timescale of seconds can be observed.[1,13,27] Jumps in PL intensity to less than half (or more than double) of its former value were observed in approximately 25% of the recorded transients. These jumps are distributed arbitrarily in time and are caused by reversible photoinduced radical formation, or by subsequent irreversible photo-oxidation.[14,28] For each transient (56 chains were measured in total, which show a suitable jump in intensity), one interval for both high and low PL intensity was selected, as marked in red and blue in the example in Fig. 2(a). As for the contacted samples described above, a correlation measurement was conducted for both measurement intervals in order to determine the photon statistics in fluorescence. Fig. 2(b) shows the fluorescence photon cross-correlation histograms of the regions marked in Fig. 2(a). The expected value of g$^{(2)}$(0) for a perfect single photon emitter is close to zero because of the signal-to-background ratio of >100 in the case of the single-molecule sample without the charge injection device. Both cases, quenched and unquenched emission, exhibit a clear dip at zero time delay. From such a histogram the ratio $N_C/N_L$ was derived with $N_C$ and $N_L$ corresponding to the number of correlation events for zero time delay (the central dip) and an arbitrary time delay between both detectors (the lateral correlation peaks), respectively. For each single molecule, two $N_C/N_L$ values were extracted, one for a region of high PL intensity and one for low PL intensity. These values are correlated in a scatter plot, shown in Fig. 2(c). As for the case of controlled carrier injection in Fig. 1, the scatter demonstrates that the $N_C/N_L$ ratio is larger for most molecules (80 %) for low PL intensities compared to high PL intensities, implying more classical photon emission in the quenched case. In addition to the correlation analysis, $\tau_{PL}$ values were determined for the same time intervals. Fig. 2(d) reports a similar trend as for the molecules in the device structure in Fig. 1(d): $\tau_{PL}$ decreases substantially with the introduction of a quencher. However, without the device structure, $\tau_{PL}$ values of 1ns and above are measured for the unquenched state. These values are not present within the device structure (see Fig. 1d) because the hole-injection layer directly impacts the PL characteristics even under a bias of -3V. The data can be rationalized in the following way. The spectral blue-shift upon charge injection suggests that formation of the quencher occurs predominantly, but not always, on low-energy chromophores.[29] The quenching of the low-energy chromophores also over-compensates a possible red-shift in emission of the remaining chromophores due to the Stark effect, which was shown to be in the order of 2nm for an electric field of 300kV/cm.[11] The electric field of a single charge at a distance of 5-10nm is approx. 150kV/cm, hence the Stark effect on the emission characteristics must be secondary. The shortening of $\tau_{PL}$ upon charge injection along with the reduction in overall brightness suggests that the charge impacts the remaining emitting higher-energy chromophores by opening a non-radiative decay channel since a charge effectively induces a further absorption band in the molecule. The change of the photon statistics upon charge injection is particularly interesting. It implies that prior to charge injection most of the energy is funneled towards a few low-energy chromophores and that charge injection predominantly occurs into one of these chromophores. Under photoinduced charge generation, it is the emitting and not the absorbing chromophore which harbors the charge. The sketch in Fig. 3 illustrates how this process will impact the photon statistics. A simplified conjugated polymer chain is depicted with three independent chromophores of different brightness and emission spectrum. Fig. 3(a) documents the case of a neutral (unquenched) chain. Here, most of the excitation energy is funneled to the low-energy chromophore (red), leading to 90% of the PL arising from the red chromophore and only 10% from both blue chromophores. The photon statistics are therefore mainly dominated by a single chromophore and are expected to be highly non-classical, i.e. showing strong photon-antibunching.[30] Upon charge injection, the low-energy chromophore is switched off as sketched in Fig. 3(b), and the remaining PL is generated equally by both blue chromophores. In this case it is expected that the fidelity of photon antibunching deteriorates under quenching and the ratio $N_C/N_L$ approaches 0.5.[22] For multiple active chromophores on the polymer chain, the distribution of photon arrival times would then tend to the classical limit. Previous measurements based on excitation and emission polarization also suggest that deactivation of some chromophores in the chain reduces the overall energy transfer efficiency between the remaining chromophores. This effect would further reduce singlet-singlet annihilation and decrease the fidelity of photon antibunching.[31]

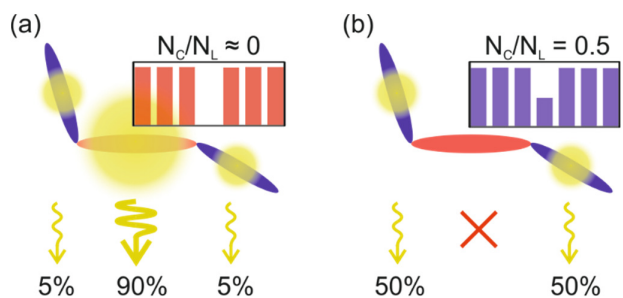

FIG. 3. A schematic of a single conjugated polymer chain consisting of three independent chromophores. (a) In the case of an unquenched chain efficient energy funneling can lead to a single dominating chromophore (red), which is responsible for 90% of the overall PL emission. The photon statistics will show strong antibunching. (b) Charge injection in the red chromophore leads to quenching of the dominant chromophore. The remaining PL stems from two equally contributing chromophores (blue). In this case, the photon statistics will show reduced photon antibunching as compared to the case in (a).

In conclusion, we have shown how the photophysics of single conjugated polymer chains changes upon charging. Our results suggest that following charge injection – either stochastically by random radical formation or deterministically by controlled charge injection in a device architecture – the charge is situated on a low-energy chromophore, quenching its PL, which leads to a slight overall blue-shift of the remaining PL emission. Additionally, the PL lifetime, $\tau_{PL}$, is decreased, due to the impact of the charge on the remaining chromophores. This quenching leads to the counter-intuitive behavior with regard to fluorescence photon statistics: one would expect quenching to reduce the number of chromophores and therefore to improve the fidelity of antibunching, but the opposite is true, both under stochastic and under deterministic charging. The photon statistics of a multi-chromophoric object, like a conjugated polymer chain, can in principle be changed reversibly by charge injection. However, the energetic landscape, responsible for the energy funneling, plays a major part in this quenching and has to be taken into account for the correct interpretation of photon statistics. Turning off the fluorescence of the lowest-energy chromophore can lead to dominating emission from the remaining chromophores within the chain, while fluorescence quenching is observed.

Acknowledgements

The authors thank Dr. Takuji Adachi and Florian Steiner for assistance in device fabrication, and Prof. Dieter Weiss for providing access to device fabrication facilities. The authors are indebted to the European Research Council for funding through the Starting Grant MolMesON (No. 305020).